\documentclass[pra,twocolumn,fleqn]{revtex4-1}
\usepackage{graphicx}
\usepackage{amsmath}
\usepackage{amssymb}
\usepackage{amsmath}
\usepackage{color}
\usepackage{bm}
\usepackage{latexsym}
\usepackage{hyperref}
\usepackage{float}
\usepackage{graphicx}
\usepackage{epstopdf}
\usepackage{subfigure}
\newcommand{\void}[1]{}

\newcommand{\be}{\begin{equation}}
\newcommand{\ee}{\end{equation}}

\begin{document}
\title{Simulating spin measurement with a finite heat bath model for the  environment}
\author{Thomas Dittrich}
\affiliation{Departamento de F\'\i sica, Universidad Nacional de Colombia, Bogot\'a, Colombia}
\author{Oscar Rodr\'\i guez}
\affiliation{Departamento de F\'\i sica, Universidad Nacional de Colombia, Bogot\'a, Colombia}
\author{Carlos Viviescas}
\affiliation{Departamento de F\'\i sica, Universidad Nacional de Colombia, Bogot\'a, Colombia}
\date{\today}

\date{\today}

\begin{abstract}
Spin measurement is studied as a unitary time evolution of the spin coupled to an environment representing the meter and the apparatus. Modelling the environment as a heat bath comprising only a finite number of boson modes and represented in a basis of coherent states, following the Davydov ansatz, it can be fully included in the quantum time evolution of the total system. We perform numerical simulations of projective measurements of the polarization, with the spins prepared initially in a neutral pure state. The likewise pure initial state of the environment is constructed as a product of coherent states of the boson modes with a random distribution of their centroids around the origin of phase space. Switching the self-energy of the spin and the coupling to the heat bath on and off by a time-dependent modulation, we observe the outcome of the measurement in terms of the long-time behaviour of the spin. Interacting with the heat bath, the spins get entangled with it and lose coherence, thus reproduce the ``collapse of the wavefunction''. The expected quantum randomness in the final state is manifest in our simulations as a tendency of the spin to approach either one of the two eigenstates of the measured spin operator, recovering an almost pure state.  The unitary time evolution allows us to reproducibly relate these random final states to the respective initial states of the environment and to monitor the exchange of information between the two subsystems in terms of their purity and mutual entropy.
\end{abstract}

\maketitle

\section{Introduction}

Since the Copenhagen interpretation, measurement has played a special role in quantum mechanics. It is indispensable as a link to macroscopic classical processes, but for this reason does not fit seamlessly in the formal framework of quantum mechanics proper. The Copenhagen interpretation postulates in particular the collapse of the wave function as an irreversible discontinuous jump from coherent superpositions to alternative classical facts. This ambiguous function has elicited a debate on all aspects of quantum measurement that lasts till today.

The first systematic mathematical account of observation in quantum mechanics, owed to von Neumann \cite{Neu18}, lists a sequence of steps that are essential for what is called a projective quantum measurement. Two of them appear incompatible with the unitary time evolution generated by the Schr\"odinger equation: (i), the reduction of the density operator of the measured system to a set of probabilities for the eigenvalues of the measured observable to be actually observed (the ``first'' collapse of the wave function) and (ii), the projection of the state of the measured system onto that eigenstate of the measured observable corresponding to the observed eigenvalue (sometimes referred to as the ``second collapse''), to warrant the immediate repeatability of the measurement.

Of these two steps, the first collapse could be reconciled with the unitary framework of quantum time evolution as a manifestation of decoherence. Adopting the microscopic approach to decoherence and dissipation pioneered by Feynman and Vernon \cite{FV63}, Zurek \cite{Zur81,Zur84}, Joos and Zeh \cite{JZ85}, and others developed microscopic models for the loss of coherence in the measured system, coupling it to a large number of degrees of freedom of the measurement apparatus and its environment. In this way, the collapse not only became amenable to a detailed physical analysis, but in particular could be resolved as a gradual process taking place in continuous time \cite{Zur81,JZ85,Ven97}. The progress in the understanding of quantum measurement achieved by the decoherence approach is undeniable. It is an essential step towards integrating the measurement process in an overall unitary description. However, to accomplish the irreversible collapse for the part of object, it has to project out the degrees of freedom of the apparatus and treat the measured system in a statistical fashion, in terms of its reduced density operator evolving in time according to master equations, Redfield equations, and similar schemes. In this way, it foregoes access to individual runs of the measurement and obviously to the possibility of resolving any internal structures of the macroscopic apparatus. Evidently, the second collapse is out of reach for the decoherence approach.

An explicit unitary account of the entire measurement setup seems unfeasible, either, because in order to reach full decoherence in the first collapse, it is necessary to let the number $N$ of modes of the environment approach infinity. There are other, subtler but no less cogent, arguments against a unitary treatment, one of them asserting that the second collapse is incompatible with the principle of coherent superposition, unconditionally valid in a unitary framework \cite{Sto16}. We here explore a hybrid perspective that attempts to take advantage of the virtues of a unitary treatment, but surmounts its main deficiency, the inability to reproduce the first collapse. As a decisive improvement, we combine the unitary approach with a method recently proposed for quantum optics, quantum chemistry, mesoscopic physics and related fields: The finite heat bath method \cite{GKG10,GBS14} has been introduced as an alternative, in particular in numerical simulations, to the traditional modelling of decoherence and dissipation based on the incoherent dynamics of the reduced density operator of the central system. Instead, it takes all the modes of the heat bath explicitly into account, restricting their number to a large but finite value $N \gg 1$. A system with a dense but still discrete spectrum evolving unitarily should invariably show revivals of coherence on long time scales. Indeed, a detailed model of a quantum optical high Q cavity coupled to another cavity playing the role of the environment \cite{RBH97} suggests that an apparent decoherence of Schr\"odinger cats in the first cavity is followed by sharp revivals on longer time scales.

Yet numerical experiments with finite heat baths \cite{GKG10,GBS14} provided ample evidence that even with moderate values of $N$, systems converge to full decoherence on all practically relevant time scales. Applying the finite heat bath method to quantum measurement not only allows us to reproduce the collapse, but to accomplish much more: We can define all details of the initial states of the measured object \emph{and} the environment for each run and calculate the time evolution of the entire system reproducibly for each initial condition through the full measurement process. This permits for example to simulate measurements with linear superpositions of particular initial states of the environment and thus to check the superposition argument experimentally. The fact that in a closed system, the total quantum entropy is conserved \cite{Weh91,Dit19} opens the possibility to compare the entropy balances of measurement object and environment and to monitor the sharing and exchange of information between them. Surprisingly, we are even able to reproduce essential features of the second collapse.

In this paper, we work out our approach for the case of spin measurement. Determining the polarization of spin-$1/2$ particles is a prototypical instance of quantum measurement \cite{Zei99}. It combines simplicity with exceptional accessibility to analytical calculations. In particular, spin measurement is a paradigm of quantum randomness. Our approach results in a kind of spin-boson model with a finite number of boson modes, which we use as a starting point for extensive numerical simulations of spin measurements. We focus on measurements of $\hat\sigma_z$ on spins prepared in an eigenstate of $\hat\sigma_x$ with random initial conditions of the apparatus and analyze the statistics of the states approached for long times. The most important result is that the spins, after losing all coherence, recover nearly pure states again. However, instead of the discrete alternative, spin up or down, expected for a projective measurement, we find a continuous bimodal distribution, but with a preference for the two extremes. The random scattering of the individual outcomes can be traced back to the randomness of the initial states of the environment.

We construct our model, discuss its symmetries, and compare it with other categories within the broad variety of spin-boson models in Section \ref{sec21}. Of vital importance for our approach is the precise modelling of the spectrum of the boson sector, a crucial factor for the dynamical behaviour of the spin, detailed in Section \ref{sec22}. Section \ref{sec3} is dedicated to our numerical experiments. We explain their protocol in Section \ref{sec31}, in particular the preparation of the initial states of spin and heat bath, a decisive issue for our approach (Section \ref{sec32}). The results of the simulations are then presented in Section \ref{sec33}, focussing on the time evolution of the spin and the character and statistics of its long-time behaviour (\ref{sec331}), the interchange of information between spin and environment in the course of the measurement (\ref{sec332}), and special situations that arise if the spin is already prepared in an eigenstate of the measured observable or in repeated measurements (\ref{sec333}). In Subsection \ref{sec334}, we address simulations of measurements with the environment prepared in linear superpositions of specific initial states, a key issue in the debate on the unitary approach to quantum measurement. We conclude in Section \ref{sec4} with a resume what we have achieve with a unitary approach to spin measurement combined with the finite heat bath method, and an outlook to pertinent issues left open.

\section{\label{sec2} Modelling spin measurement with finite heat baths}

\subsection{\label{sec21} Constructing the model}

Most of the literature on quantum measurement assumes a partition of the total setup into three principal components, the measured object (``O''), the apparatus or meter (``M''), possibly still of microscopic nature, that directly couples to and gets entangled with the object, and a macroscopic part, the environment (``E''), that converts the quantum information on the object shared by the meter into classically observable facts, inducing a loss of coherence in the object and meter \cite{VKG95}. This tripartite structure would correspond to a Hamiltonian for the entire system comprising five terms, the self-energies of the three components and the couplings between object and meter and between meter and environment,
\begin{equation} \label{omeham}
\hat{H} = \hat{H}_{\rm O} + \hat{H}_{\rm OM} + \hat{H}_{\rm M} + \hat{H}_{\rm ME} + \hat{H}_{\rm E}.
\end{equation}
For the objectives of the present work, the internal structure of the meter and the coherent process of its entanglement with the object are not of primary importance. In order to simplify the model as far as possible, we therefore merge meter and environment, reducing the number of components to two and the Hamiltonian to three terms, the self-energies of object and environment and their coupling,
\begin{equation} \label{ooeeham}
\hat{H} = \hat{H}_{\rm O} + \hat{H}_{\rm OE} + \hat{H}_{\rm E}.
\end{equation}
We assume a standard energy splitting $\hbar\omega_0$ for the spin, so that
\begin{equation} \label{oham}
\hat{H}_{\rm O} =  \frac{1}{2} \hbar\omega_0\hat\sigma_x
\end{equation}
(adopting standard notation for  the Pauli operator $\hat\sigma_x$). In the context of spin measurement, the self-energy term represents the preparation of the spin in a specific initial state, prior to the measurement proper. It is typically accomplished by aligning the spin to a magnetic field $\mathbf{B} = B \mathbf{u}_B$ in the direction $\mathbf{u}_B$ of the intended initial polarization, here the $x$-direction, related to the energy splitting by $\hbar\omega_0 = \mu B$ (with $\mu = g_s \mu_{\rm{B}} / 2$, the magnetic moment of the spin, $g_s$ denoting the spin $g$-factor and $\mu_{\rm{B}}$ the Bohr magneton).

We model the environment as a set of $N$ harmonic oscillators with discrete frequencies $\omega_n$, $n = 1,\,\ldots,\,N$, to be specified further below,
\begin{equation} \label{eham}
\hat{H}_{\rm E} =  \sum_{n=1}^N \hbar\omega_n \left(\hat a_n^\dagger \hat a_n + \frac{1}{2}\right).
\end{equation}
The interaction between spin and environment, now interpreted as the meter, is chosen as a linear coupling between the spin operator $\hat\sigma_z$ and the position operators $\hat a_n^\dagger + \hat a_n$ of the bath modes, with individual coupling constants $g_n$ \cite{DG90C}
\begin{equation} \label{oeham}
\hat{H}_{\rm OE} = \sum_{n=1}^N  g_n \hat\sigma_z (\hat a_n^\dagger + \hat a_n).
\end{equation}
With this choice, $\hat\sigma_z$ assumes the role of the measured observable, so that we are modelling measurements of the $z$-component of the spin or its vertical polarization. It is crucial that the spin operator in the coupling term does not coincide with that in the self-energy (\ref{oham}) of the spin, otherwise the energy splitting would impose a bias on the measurement.

The total Hamiltonian thus takes the form of a spin-boson model,
\begin{align} \label{spinbosonham}
\hat{H} = & \frac{1}{2} \hbar\omega_0\hat\sigma_x f_{\rm O}(t) + \nonumber
\sum_{n=1}^N  g_n \hat\sigma_z (\hat a_n^\dagger + \hat a_n)\, f_{\rm OE}(t) + \\
& + \sum_{n=1}^N \hbar\omega_n \left(\hat a_n^\dagger \hat a_n + \frac{1}{2}\right).
\end{align}
The time-dependent modulation functions $f_{\rm O}(t)$ and  $f_{\rm OE}(t)$ allow us a more detailed modelling of the measurement protocol, for example choosing sigmoid or box-shaped profiles to switch them on or off. A two-level atom in a high Q microwave cavity is one of the numerous possible experimental realizations of this model.

The spin-boson Hamiltonian is an important workhorse of theoretical quantum optics and atomic physics and has been intensely studied for various regimes of the boson number $N$. The case $N = 1$, known as quantum Rabi model \cite{BC&16}, is a standard model for light-matter interaction. Powerful approximations such as the rotating-wave approximation \cite{GZ00} allow to explore the Rabi model in wide parameter regimes. Even without such approximations, the spectrum can be solved analytically \cite{Bra11,XZ&17}. In the opposite limit $N \to \infty$, the spin-boson Hamiltonian is a prototype of dissipative two-state systems. Its long-time behaviour is known for various types of the heat-bath spectrum \cite{LC&87}. The regime of moderate values of $N$ addressed here, on the other hand, is hardly explored till now. As we are interested in coming close to ``genuine'' decoherence as in systems with continuous spectrum, however, the limit $N \to \infty$ is a pertinent reference and benchmark for us.

\begin{figure}[h!]
\begin{center}
\includegraphics[width=8.6cm]{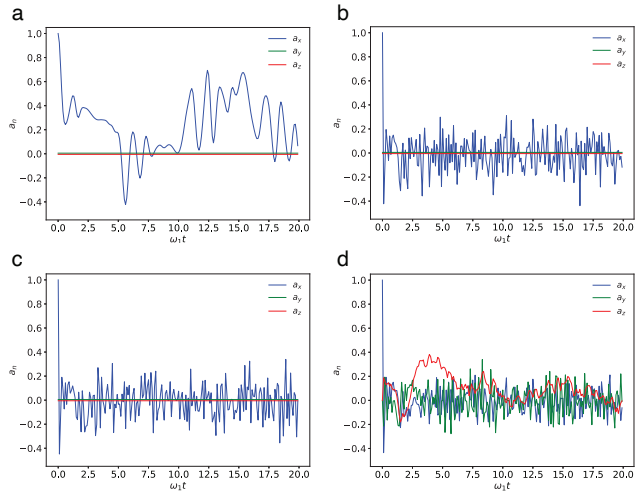}
\caption{Simulation of spin measurements under well-defined symmetry conditions concerning the parity $P_z:\; z \to -z$, Eq.\ (\protect\ref{spinmetsym}). Panels show the time evolution of the components $a_x$ (blue), $a_y$ (green), $a_z$ (red), of the measured spin, prepared initially in the symmetric state $| \psi_{\rm{O}}(0) \rangle = |+_x \rangle$ (element of the symmetric subspace $\mathcal{H}_{{\rm O}+}$), as functions of scaled time $\omega_1 t$ under the Hamiltonian (\protect\ref{spinbosonham}) with a single boson mode, $N = 1$ (Rabi model), with particular initial states of the harmonic oscillator as follows: (a) ground state $|0_1 \rangle$, (b) superposition of even Fock states $|(2n)_1 \rangle$, $n \in \mathbb{Z}$, (element of the symmetric subspace $\mathcal{H}_{{\rm E}+}$), (c) superposition of odd Fock states $|2(n+1)_1 \rangle$ (element of the antisymmetric subspace $\mathcal{H}_{{\rm E}-}$), (d) superposition of even and odd Fock states (no specific parity). Parameter values are $\omega_0 / \omega_1 = 4$ and $g_1 / \omega_0 = 2$.
}
\label{fig1_symmetry}
\end{center}
\end{figure}

An important feature of the Hamiltonian (\ref{spinbosonham}) is its invariance under the reflection (parity) $P_z:\; z \to -z$. The inversion operators $\hat\Pi_{z,{\rm O}} = \hat\sigma_x$ for the spin and $\hat\Pi_{z,{\rm E}} = \exp\left({\rm i}\pi\sum_n \hat a_n^\dagger \hat a_n\right)$ \cite{Bru07} for the operators of the bath combine to a total parity operator
\begin{equation} \label{spinmetsym}
\hat\Pi_z = \hat\Pi_{z,{\rm O}}\hat\Pi_{z,{\rm E}} =
\hat\sigma_x \exp\left({\rm i}\pi\sum_n \hat a_n^\dagger \hat a_n\right).
\end{equation}
 The invariance $\hat\Pi_z^\dagger \hat H\hat\Pi_z = \hat H$ of the total Hamiltonian is readily verified. With this symmetry, the Hilbert space $\mathcal{H}$ of the total system decomposes into two eigensubspaces of $\hat\Pi_z$,
\begin{equation} \label{paritydecomp}
\mathcal{H} = \mathcal{H}_+ \otimes \mathcal{H}_-.
\end{equation}
The Hilbert spaces of spin and bath, in turn, decompose each into their even and odd subspaces $\mathcal{H}_{{\rm O}+}$,  $\mathcal{H}_{{\rm O}-}$ and $\mathcal{H}_{{\rm E}+}$,  $\mathcal{H}_{{\rm E}-}$, resp., under $P_z$: For the spin, they are spanned by the eigenstates of $\hat\sigma_z$, $\hat\sigma_z \left|{z_\pm}\right\rangle = \pm \left|{z_\pm}\right\rangle$, and for each bath mode by its even and odd energy eigenstates, resp. In this way, the even (symmetric) and odd (antiymmetric) subspaces of $\mathcal{H}$ can be broken down further into
\begin{equation} \label{sbparitydecomp}
\begin{split}
\mathcal{H}_+ &= \mathcal{H}_{{\rm O}+} \otimes \mathcal{H}_{{\rm E}+} \oplus
\mathcal{H}_{{\rm O}-} \otimes \mathcal{H}_{{\rm E},-}, \\
\mathcal{H}_- &= \mathcal{H}_{{\rm O}+} \otimes \mathcal{H}_{{\rm E}-} \oplus
\mathcal{H}_{{\rm O}-} \otimes \mathcal{H}_{{\rm E}+}.
\end{split}
\end{equation}
This symmetry has crucial consequences for spin measurement: If the initial state of the total system belongs to one of the invariant subspaces $\mathcal{H}_+$ or $\mathcal{H}_-$, it can evolve under $\hat{H}$ only into states of the same symmetry class with respect to $P_z$. If in addition, the spin is prepared before the measurement in a pure state with $\langle \hat\sigma_z \rangle = 0$, for example an eigenstate of $\hat\sigma_x$ or $\hat\sigma_y$, it is itself symmetric under $P_z$ and belongs to $\mathcal{H}_{{\rm O},+}$ or $\mathcal{H}_{{\rm O},-}$. In order that after the measurement, the spin exits as a polarized state, in particular as $\left|{z_+}\right\rangle$ or $\left|{z_-}\right\rangle$, the total initial state must not have been an element of $\mathcal{H}_+ \oplus \mathcal{H}_-$. This is only possible if the initial state \emph{of the bath}, in turn, had violated $P_z$. A numerical example of this conditionality is given in Fig.\ \ref{fig1_symmetry}. From the point of view of invariance under $P_z$, a symmetry breaking by the initial condition of meter and apparatus is therefore a necessary condition for a definite outcome of the measurement, spin up or down, if the initial state of the spin itself does not have such a bias. Since the invariant subspaces $\mathcal{H}_{{\rm O},+}$ and $\mathcal{H}_{{\rm O},-}$ are of measure zero within $\mathcal{H}_{\rm O}$, an initial state of the bath, chosen at random, will break the symmetry anyway with probability 1. This is the most relevant situation for quantum randomness and most challenging for our approach, we shall therefore focus our simulations on the particular case of the spin being prepared in a symmetric initial state, while the initial condition of the bath is defined by a probability distribution that is symmetric on average, but can break the symmetry by random fluctuations of the modes composing it. For the same reason, we abstain from considering random initial conditions also for the spin, as they would arise in a traditional Stern-Gerlach experiment \cite{GS22a,GS22b}.

\subsection{\label{sec22} Spectrum of the heat bath}

For models of dissipation and decoherence with continuous spectrum of the heat bath, the spectral distribution is a crucial ingredient. It determines the influence of the bath on the central system and thus, for example, which type of damping it induces (Ohmic, subohmic, superohmic  \cite{LC&87}). In our case of a discrete spectrum with a large number of frequencies, it is appropriate to define the frequencies in terms of a continuous spectral distribution, adopting the types of spectra common in continuous models, and then to discretize it in a suitable manner. The fundamental quantity defining the spectral distribution is the frequency density $\rho_\omega(\omega)$, for a discrete spectrum given by
\begin{equation} \label{freqdens}
\rho_{\rm{f}}(\omega) =
\sum_{\scriptstyle n \atop \scriptstyle \omega \le \omega_n \le \omega + {\rm{d}} \omega}
\delta(\omega - \omega_n).
\end{equation}
Where it is adequate to determine the spectrum directly in terms of the frequency density, a desired distribution can be achieved by fixing the level separations accordingly \cite{DP20}. In our case, the impact of the environment on the measured object is given by the interaction term, Eq.\ (\ref{oeham}), and depends as much on the coupling constants $g_n$ as on the spectrum of the environment itself. In this case, it is more appropriate to define the spectrum in terms of the spectral density \cite{LC&87,HW&19}
\begin{equation} \label{specdens}
J(\omega) = \pi \sum_n g_n^2 \delta(\omega - \omega_n).
\end{equation}
Instead of modulating the level spacings, a desired spectral density $J(\omega)$ can now be achieved by modulating the coupling constants accordingly, while the frequency density can be any convenient normalized function, $\int_0^\infty {\rm{d}} \omega \,\rho_{\rm{f}}(\omega) = 1$ \cite{HW&19}.

A standard form of the spectral density for continuous spectra combines an algebraic increase $\sim \omega^s$ for low frequencies with an exponential high-frequency cutoff at $\omega_{\rm{c}}$ \cite{LC&87,HW&19},
\begin{equation} \label{ohmdens}
J(\omega) = 2 \pi \alpha \, \omega_{\rm{c}}^{1-s} \omega^s {\rm{e}}^{-\omega / \omega_{\rm{c}}}.
\end{equation}
The Kondo parameter $\alpha$ is a dimensionless global measure of the system-environment coupling. Spectra are classified according to the exponent $s$ as sub-Ohmic ($s < 1$), Ohmic ($s = 1$), or super-Ohmic ($s > 1$). To our best knowledge, established models for the environment or the apparatus in the context of quantum measurement, with specific physically plausible assumptions for the spectral density, do not exist. In this work, we therefore vary $s$ as well as $\alpha$ to identify parameter regimes that appear suitable for a realistic account of spin measurement.

\section{\label{sec3} Numerical experiments}

\subsection{\label{sec31} Methods}
Simulating the time evolution of the spin-boson system requires to solve the time-dependent Schr\"odinger equation with the Hamiltonian (\ref{spinbosonham}). A straightforward approach would be representing this Hamiltonian in the direct product basis of energy eigenstates $|\sigma_x\rangle \bigotimes_{n = 1}^N | j_n\rangle$
of spin and bath modes,
\begin{align} \label{dirprod}
\hat{H}_{\rm O} |\pm \rangle &= \frac{\sigma_x}{2} \hbar\omega_0 |\sigma_x \rangle, \; \sigma_x = \pm 1, \nonumber \\
\hat{H}_{\rm E} \bigotimes_{n = 1}^N | j_n\rangle  &= \sum_{n=1}^N \left(j_n + \frac{1}{2}\right) \hbar\omega_n | j_n\rangle, \;
j_n \in \mathbb{N}_0 ,
\end{align}
diagonalizing $\hat{H}$ in this basis, and propagating arbitrary initial states in the basis of its eigenstates. However, the exponential increase with the number of bath modes of the memory space required to accommodate the Hamiltonian matrix is prohibitive.

A more promising approach is to start from a basis for the bath modes adapted to the problem, according to the following criteria: Basis states (i) are readily accessible, (ii) are at least not too different from the expected eigenstates of the Hamiltonian, (iii) are close to or coincide with the intended initial states of the simulations, and (iv) evolve in time in a way that is relatively easy to compute. In the case of the spin-boson model, it is in particular coherent states of the bath oscillators that meet these conditions. For a single harmonic oscillator with ground state $|0\rangle$, a general coherent state is defined as
\begin{equation} \label{cohstat}
| \gamma\rangle = \hat D(\gamma) |0\rangle ,
\end{equation}
where
\begin{equation} \label{dispop}
\hat D(\gamma) = \exp\left(\gamma \hat a^\dagger - \gamma^\ast \hat a \right)
\end{equation}
is a displacement operator, and the complex parameter $\gamma = \sqrt{m\omega/2\hbar} \bigl(q + ({\rm{i}}p/m\omega)\bigr)$ determines the phase-space coordinates ${\mathbf{r}} = (p,q)$ of the centroid of the coherent state.

Since the overlap of two coherent states $| \gamma\rangle$, $| \gamma' \rangle$,
\begin{equation} \label{gamma}
\big | \langle \gamma | \gamma^\prime \rangle \big |^2 = \exp\left(- |\gamma^\prime - \gamma |^2 \right),
\end{equation}
is always positive, they cannot form an orthonormal basis. However, discrete subsets of the continuous family of coherent states can be over- as well as incomplete. Placing the centroids on a square grid such that each cell occupies exactly a Planck cell in phase space, forming a von Neumann lattice, results in an exactly complete basis \cite{Neu18,BB&71,Per71}. In order to apply coherent states to the integration of the time-dependent Schr\"odinger equation, a time dependent position of their centroids is readily introduced as 
\begin{equation} \label{tdcohstat}
\begin{split}
| \gamma(t)\rangle &= \hat D\bigl(\gamma(t)\bigr) |0\rangle , \\
\hat D\bigl(\gamma(t)\bigr) &= \exp\left(\gamma(t)  \hat a^\dagger - \gamma^\ast (t) \hat a \right), \\
\gamma(t) &= \sqrt{m\omega/2\hbar} \bigl(q(t) + {\rm{i}}p(t)/m\omega \bigr).
\end{split}
\end{equation}

We here adopt an implementation of the approximate solution of the time-dependent Schr\"odinger equation based on coherent states, the Davydov Ansatz \cite{DK73,Dav73,Dav80,Dav85}, that is particularly suited for our purposes. It introduces time-dependent expansion coefficients for the coherent states representing the bath modes as well as for the basis states of the central system, i.e. here, the spin,
\begin{align} \label{d2ansatz}
| \Psi_{\rm{D2}}(t)\rangle &= \bigl(C_+(t) | + \rangle + C_-(t) | - \rangle \bigr)
|{\boldsymbol{\gamma}}(t) \rangle , \nonumber \\
|{\boldsymbol{\gamma}}(t) \rangle &= \bigotimes_{n = 1}^N
| \gamma_n(t) \rangle ,
\end{align}
with $| \gamma_n(t) \rangle$ defined as in Eq.\ (\ref{tdcohstat}). 
In this form, it is known as the D2-Ansatz. Variations and refinements are possible \cite{WG18}, for example, different coherent state parameters $\gamma_n^+(t)$, $\gamma_n^-(t)$ can be assumed for the two spin terms in Eq.\ (\ref{d2ansatz}) (so-called D1-Ansatz). Since the convergence towards stable long-time states depends sensitively on the number of nearly orthogonal coherent states included in the Davydov ansatz, we increase their quantity by a computationally cost-efficient enhancement of Eq.\ (\ref{d2ansatz}), known as multimode ansatz,
\begin{align} \label{d2multi}
| \Psi_{\rm{D2}}(t)\rangle &= \sum_{m = 1}^M \bigl(C_{m+}(t) | + \rangle + C_{m-}(t) | - \rangle \bigr)
|{\boldsymbol{\gamma}} _m(t)\rangle, \nonumber \\
|{\boldsymbol{\gamma}}_m(t) \rangle &= \bigotimes_{n = 1}^N
| \gamma_{mn}(t) \rangle .
\end{align}
It introduces independent amplitudes $C_{m\pm}(t)$ of the spin states for each one of a set of $M$ configurations $|{\boldsymbol{\gamma}}_m(0) \rangle$, $m = 1,\,\ldots,\,M$, of the $N$ coherent states of the boson modes. Distinct initial conditions $|{\boldsymbol{\gamma}}_m(0) \rangle$ are generated, for example, by shifting a given set of coherent states, say $|{\boldsymbol{\gamma}}_1(0) \rangle$, rigidly in different directions in their $2N$-dimensional common phase space. For $M > 1$, Eq.\ (\ref{d2multi}) implies substantial entanglement of object and environment. In order to make sure that the two subsystems nevertheless be uncorrelated before the measurement, see Eq.\ (\ref{oeini}) below, we have to choose the initial values of the amplitudes $C_{m\pm}(t)$ accordingly, for example as
\begin{equation} \label{cpmini}
C_{m\pm}(0) = \begin{cases}
C_{1\pm}(0) \neq 0& m = 1, \\
0 & m \neq 1.
\end{cases}
\end{equation}
Equation (\ref{d2multi}) already supposes implicitly that each bath oscillator is initiated in a single coherent state $|\gamma_{mn}(0)  \rangle = \hat D\bigl(\gamma_{mn}(0) \bigr) | 0_n \rangle$. This assumption is not only physically plausible, it is in particular an enormous  advantage for the further integration process. Substituted in the Schr\"odinger equation, Eq.\  (\ref{d2multi}) leads to evolution equations for the functions $\gamma_{mn}(t)$, which in turn take the form of Lagrangian equations of motion and can be solved with corresponding numerical methods \cite{GKG10}.

Since coherent states are not eigenstates of the spin-boson Hamiltonian, they will not remain coherent states under the time evolution and in particular lose their minimum uncertainty property. This is taken into account by including additional coherent states in the time evolution as required, selected from a set of initially unoccupied coherent states located on von Neumann lattices around the initially occupied states $\hat D\bigl(\gamma_{mn}(0) \bigr) | 0_n \rangle$. Inversely, coherent states can approach one another so closely that they become nearly degenerate. This is avoided by eliminating one state of such a pair, a programmed removal known as apoptosis \cite{WG20}.

\subsection{\label{sec32} Initial conditions}
For the present work it is essential to construct the initial state of the entire system in such a way that it represents most faithfully the conditions of a typical spin measurement. We here adopt the conventional scheme of measurements resulting in a projection of the state of the measured system onto the eigenspaces of the measured observable. Within a general formalism of quantum measurement, our model therefore belongs to the category of projective measurements \cite{NC00}. In most of the literature on the subject, it is assumed that sufficiently far before the measurement, measured object and apparatus are independent systems, i.e., the total state factorizes into two pure states,
\begin{equation} \label{oeini}
\hat\rho(0) = | \Psi(0) \rangle \langle \Psi(0) |,\quad | \Psi(0) \rangle =  | \psi_{\rm{O}}(0) \rangle | \psi_{\rm{E}}(0) \rangle.
\end{equation}
As we are modelling observations of $\hat \sigma_z$, in order to exclude an initial bias of the measurement, we have to choose $| \psi_{\rm{O}}(0) \rangle$ such that it is symmetric with respect to the parity $P_z:\; z \to -z$, i.e., located at the equator of the Bloch sphere. Eigenstates of $\hat \sigma_x$ and $\hat \sigma_y$,
\begin{equation} \label{psieini}
| \psi_{\rm{O}}(0) \rangle = |\pm_x \rangle, \;
\hat \sigma_x |\pm_x \rangle = \pm |\pm_x \rangle,
\end{equation}
(likewise for $\hat \sigma_y$) or linear combinations thereof meet this condition. For complementary simulations, see Subsection \ref{sec333}, we shall exceptionally prepare the spin also in eigenstates of $\hat \sigma_z$.

The initial condition of the environment, uncorrelated with  the central system and in a pure state not imposing a systematic bias on the measurement outcome either, should reflect the random character of the uncontrolled degrees of freedom of the apparatus. In addition, it is restricted by the rather specific structure (\ref{d2ansatz}) implied by the Davydov ansatz, that is, it should take the form of a product of coherent states $| \gamma_n \rangle$ of the individual harmonic oscillators,
\begin{equation} \label{rhoeini}
\hat\rho_{\rm{E}}(0) = | \psi_{\rm{E}}(0) \rangle \langle \psi_{\rm{E}}(0) |,\quad | \psi_{\rm{E}}(0) \rangle = \bigotimes_{n = 1}^N
| \gamma_n(0) \rangle .
\end{equation}
The simplest option, preparing all the bath modes in their ground state, i.e., $| \gamma_n(0) \rangle = |0_n \rangle$, $n = 1,\,\ldots,\,N$, owing to the symmetry of the Hamiltonian, could never result in any non-zero final polarization of spins that have been initiated likewise in a neutral state such as (\ref{oeini}). In the present context, it is essential to introduce a random component by an initial displacement of the coherent states. How to define the initial distribution $p(\gamma_n,\gamma_n^\ast)$ of their centroids appropriately, at the present exploratory stage, is largely a matter of trial and error, guided by physical inttuition. The most obvious choice is selecting them according to a Gaussian distribution centred at the origin in phase space,
\begin{equation} \label{gaussianpfunc}
p(\gamma_n,\gamma_n^\ast) = \frac{1}{\sqrt{2\pi} s} \exp\left(- \frac{|\gamma_n |^2}{2s^2} \right).
\end{equation}
If we define the variance $s^2$ as a dimensionless temperature in units of a photon energy $\hbar \omega$, $s^2 = k_{\rm{B}}T / 2 \hbar \omega$, we can interpret Eq.\ (\ref{gaussianpfunc}) as a thermal distribution of bosons, inspired by the Glauber-Sudarshan or P-function \cite{GZ00,Sch01}. It is related to the density operator as
\begin{equation} \label{rhopfunc}
\hat \rho = \int {\rm{d}}^2 \gamma \,P(\gamma,\gamma^\ast)  \,| \gamma \rangle \langle \gamma |
\end{equation}
and for a thermal state reads
\begin{equation} \label{thermalpfunc}
P(\gamma,\gamma^\ast) = \frac{1}{\pi \langle n \rangle} \exp\left(- \frac{|\gamma|^2}{\langle n \rangle} \right).
\end{equation}
With the high-temperature approximation of the mean photon number $\langle n \rangle  = 1 / (e^{\hbar \omega / k_{\rm{B}}T}-1) \to k_{\rm{B}}T / \hbar \omega$ for $k_{\rm{B}}T \gg \hbar \omega$, it takes a form resembling Eq.\ (\ref{gaussianpfunc}),
\begin{equation} \label{hightpfunc}
P(\gamma,\gamma^\ast) = \frac{\hbar \omega}{\pi  k_{\rm{B}}T} \exp\left(- \frac{\hbar \omega |\gamma|^2}{k_{\rm{B}}T} \right).
\end{equation}
A first survey of simulated spin measurements revealed however that with a simple Gaussian distribution of initial states as in Eq.\ (\ref{gaussianpfunc}), the expected convergence of the spin towards a stable long-time asymptote is very fragile. A plausible explanation is that a ``democratic'' distribution treating all modes on the same footing ignores the dominating role of the low-lying modes for the time evolution on long time scales. In order to give these modes the appropriate weight in the ensemble, it suggests itself to replace the photon frequency $\omega$, representing collectively all modes in the P-functions (\ref{thermalpfunc},\ref{hightpfunc}), by the iproper frequency $\omega_n$ of each mode, 
\begin{equation} \label{nongaussianpfunc}
P(\gamma_n,\gamma_n^\ast) = \frac{e^{\hbar \omega_n / k_{\rm{B}}T}-1}{\pi} \exp\left(- (e^{\hbar \omega_n / k_{\rm{B}}T}-1) |\gamma_n |^2 \right).
\end{equation}

\begin{figure}[h!]
\begin{center}
\includegraphics[width=8.6cm]{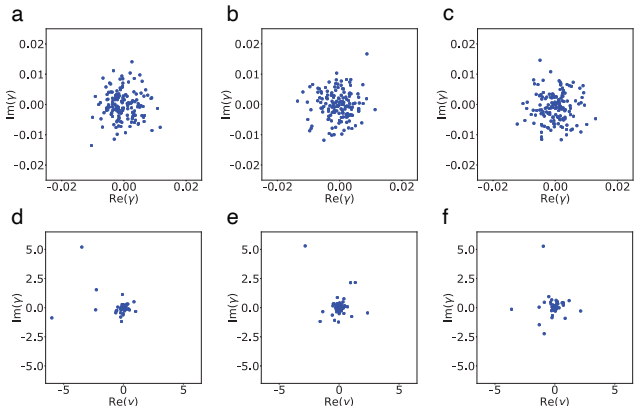}
\caption{Examples of random ensembles of initial coherent states of heat bath modes, distributed according to Eqs.\ (\protect\ref{gaussianpfunc}) (panels a, b, c) and (\protect\ref{nongaussianpfunc}) (panels d, e, f). Blue dots mark the centroids of 150 coherent states in dimensionless phase space coordinates ${\rm{Re}}(\gamma) = \sqrt{m\omega/2\hbar}\,q$, ${\rm{Im}}(\gamma) = p / \sqrt{2\hbar m\omega}$. Parameter values are $k_{\rm{B}}T = 0.2 \, \hbar\omega_0$ and $\omega_{\rm{c}}= 2 \omega_0$.
}
\label{fig2_inidis}
\end{center}
\end{figure}

In Fig.\ \ref{fig2_inidis}, we compare examples of the initial displacements of ensembles of $N = 150$ coherent states as scatter plots of their centroids in dimensionless phase space coordinates $({\rm{Re}}\,\gamma_n,{\rm{Im}}\,\gamma_n)$, generated according to the Gaussian distribution, Eq.\ (\ref{gaussianpfunc}) (panels (a) to (c)), and to the modified distribution, Eq.\ (\ref{nongaussianpfunc}) (panels (d) to (f)). It is evident that the Gaussian ensembles are more compact ,while with the modified distribution, extreme outliers, contributed predominantly by the low-lying slow modes, are significantly more frequent. It is these outliers which increase the tendency of the measured spin to converge towards eigenstates of $\hat\sigma_z$.

The initial conditions defined by Eq.\ (\ref{nongaussianpfunc}) may appear as a particular, overly restrictive choice. 
Yet, in the wider  framework of the multimode Davydov ansatz (\ref{d2multi}), they are even closed under coherent superposition and thus include superpositions of initial states of the environment, considered as a challenge for the unitary approach \cite{Sto16}. We are confident that Eq.\ (\ref{nongaussianpfunc}) represents typical states of measuring apparatuses as macroscopic many-body systems, thus granting credibility to the general conclusions we draw in Sect.\ \ref{sec4}.

\subsection{\label{sec33} Results}

\subsubsection{\label{sec331} Simulations of standard spin measurements with unbiased initial state}

The main part of our numerical experiments are simulations of measurements of $\sigma_z$ under conditions expected to lead to an unbiased random outcome for the polarization. This requires the spin to be prepared as in Eq.\ (\ref{oeini}). Initial conditions of the heat bath modes are selected at random as in Eqs.\ (\ref{rhopfunc} - \ref{hightpfunc}). For the present context, suffice itto focus on values for pertinent parameters, particularly of the coupling with the heat bath modes and their spectrum, that appear most promising for simulations of spin measurements. A more systematic survey of the parameter regimes of our model is deferred to upcoming work. Some of the simulations shown in this section have been performed with $N = 150$ bath modes, a viable number that allows a sufficiently good convergence of the spin state with the limited computing capacity available to us. Complementary calculations with $N = 300$ modes enabled significant improvements in some cases. We implemented the spectral density of these modes according to Eq.\ (\ref{ohmdens}) with $s = 0.25$, that is, our simulations are located in the deep subohmic regime of the heat bath coupling.

We implemented different measurement protocols, defining arbitrary time-dependent modulations $f_{\rm O}(t)$ of the self-energy and $f_{\rm OE}(t)$ of the coupling term in the Hamiltonian (\ref{spinbosonham}). In the sequel, we present detailed results in particular for two cases, (i) time-independent modulation functions, so that preparation and measurement take place simultaneously, and (ii) modelling a protocol where the preparation defined by $f_{\rm O}(t)$ antecedes the measurement proper, controlled by $f_{\rm OE}(t)$. We prepared the spin throughout in the neutral initial state \ref{rhoeini}, that is, with $a_x(0) = 1$ and $a_y(0) = a_z(0) = 0$. As the most specific output data, we show examples of the time evolution of the spin, represented by the reduced density operator $\bar{\hat\rho}_{\rm{O}} = {\rm{tr}}_{\rm{E}} [\hat\rho]$ and depicted as the three components of the Bloch vector $\mathbf{a}(t)$,
\begin{align} \label{rhoo}
\bar{\hat\rho}_{\rm{O}} &= a_0(t) \hat I_{\rm{O}} + \mathbf{a}(t) \cdot \hat{\boldsymbol{\sigma}}, \nonumber \\
\mathbf{a}(t) &= \bigl(a_x(t),a_y(t),a_z(t)\bigr), \; \hat{\boldsymbol{\sigma}}= \bigl(\hat\sigma_x,\hat\sigma_y,\hat\sigma_z\bigr).
 \end{align}

\begin{figure}[h!]
\begin{center}
\includegraphics[width=8.6cm]{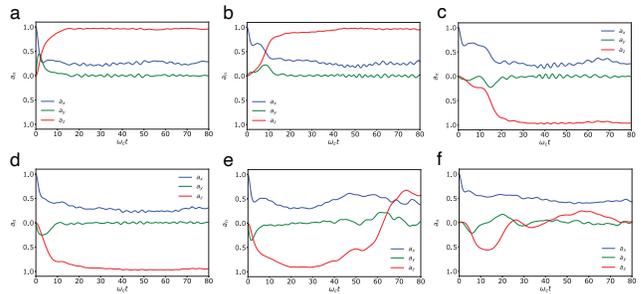}
\caption{Simulation of measurements of the polarization of the spin, with unbiased initial state $| \psi_{\rm{O}}(0) \rangle = |+ x \rangle$ of the spin and random initial conditions of meter and environment, Eqs.\ (\protect{\ref{rhopfunc}} - \protect{\ref{hightpfunc})}, modelled by the Hamiltonian (\protect{\ref{spinbosonham}}) with time-independent modulations $f_{\rm O}(t) = \rm{const}$ and  $f_{\rm OE}(t) = \rm{const}$ (see text). Panels (a) to (f) show the three components $a_x(t)$ (blue), $a_y(t)$ (green), and $a_z(t)$ (red) of the Bloch vector vs.\ time for different specific initial conditions of the environment. Time is scaled as $\omega_{\rm{c}} t$, with the cutoff frequency $\omega_{\rm{c}}$ of the heat bath spectrum, see Eq.\ (\protect{\ref{ohmdens}}). Parameter values are $N = 150$, $M = 10$, $\alpha = 0.3$, $k_{\rm{B}}T = 0.2 \, \hbar\omega_0$, $\omega_{\rm{c}}= 2 \omega_0$.
}
\label{fig3_blochvstnomodul}
\end{center}
\end{figure}

\begin{figure}[h!]
\begin{center}
\vspace{-0.5 cm}
\includegraphics[width=5cm]{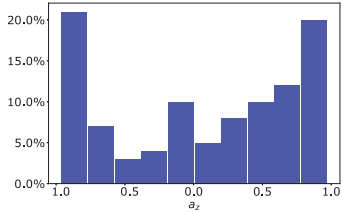}
\caption{Distribution of long-time asymptotes $a_z(t)$ for $t \gg 2\pi / \omega_{\rm{c}}$, counting only those 41 simulations among a sample of 100 runs that converge to a stable long-time average. Parameter values as in Fig.\ \protect{\ref{fig3_blochvstnomodul}}.
}
\label{fig4_asympolarnomodul}
\end{center}
\end{figure}

Figure \ref{fig3_blochvstnomodul} shows a selection of examples of the time evolution of the Bloch vector for constant modulation functions and negative splitting energy $\hbar\omega < 0$, so that the initial state of the spin coincides with its ground state. While in all cases, the $x$-component (blue) of the Bloch vector decays from its initial value 1 to a small residual level around 0.25 and the $y$-component (green) fluctuates around zero, we observe essentially three types of behaviour of $a_z(t)$ (red), the measured observable: In some cases, it approaches a positive value close to 1 and localizes there (panels a, b), in other cases it approaches a negative value close to $-1$ (panels c, d), in many instances it continues fluctuating with large amplitude around zero (panels e, f). All kinds of intermediate behaviour between these categories occur as well. Since in most cases an asymptotic value of $a_z(t)$ for $t \to \infty$ can barely be defined, a comprehensive statistical account of these asymptotes does not make sense. If we only count those simulations where a long-time average of $a_z$ can be read off with sufficient precision, some 40\% of the total sample of 100  simulations, their probability distribution appears nearly symmetric, showing a preference for the eigenvalues $a_z = -1$ and $a_z = 1$  of $\hat\sigma_z$ but also a significant lower plateau between these peaks (Fig.\ \ref{fig4_asympolarnomodul}).

This scenario changes drastically if we allow for non-trivial time dependences of $f_{\rm O}(t)$ and $f_{\rm OE}(t)$. In order to simulate a typical measurement process, we assume the following protocol: The magnitude $f_{\rm O}(t)$ of the self-energy takes a positive value initially, representing the preparation of the spin in the ground state of $\hat H_{\rm{O}}$. It is then switched off, while the coupling strength $f_{\rm O}(t)$ rises from its initial value zero to a positive maximum and then decays, following a box function to model the measurement proper, e.g., the interaction with a Stern-Gerlach magnet.

\begin{figure}[h!]
\begin{center}
\includegraphics[width=8.6cm]{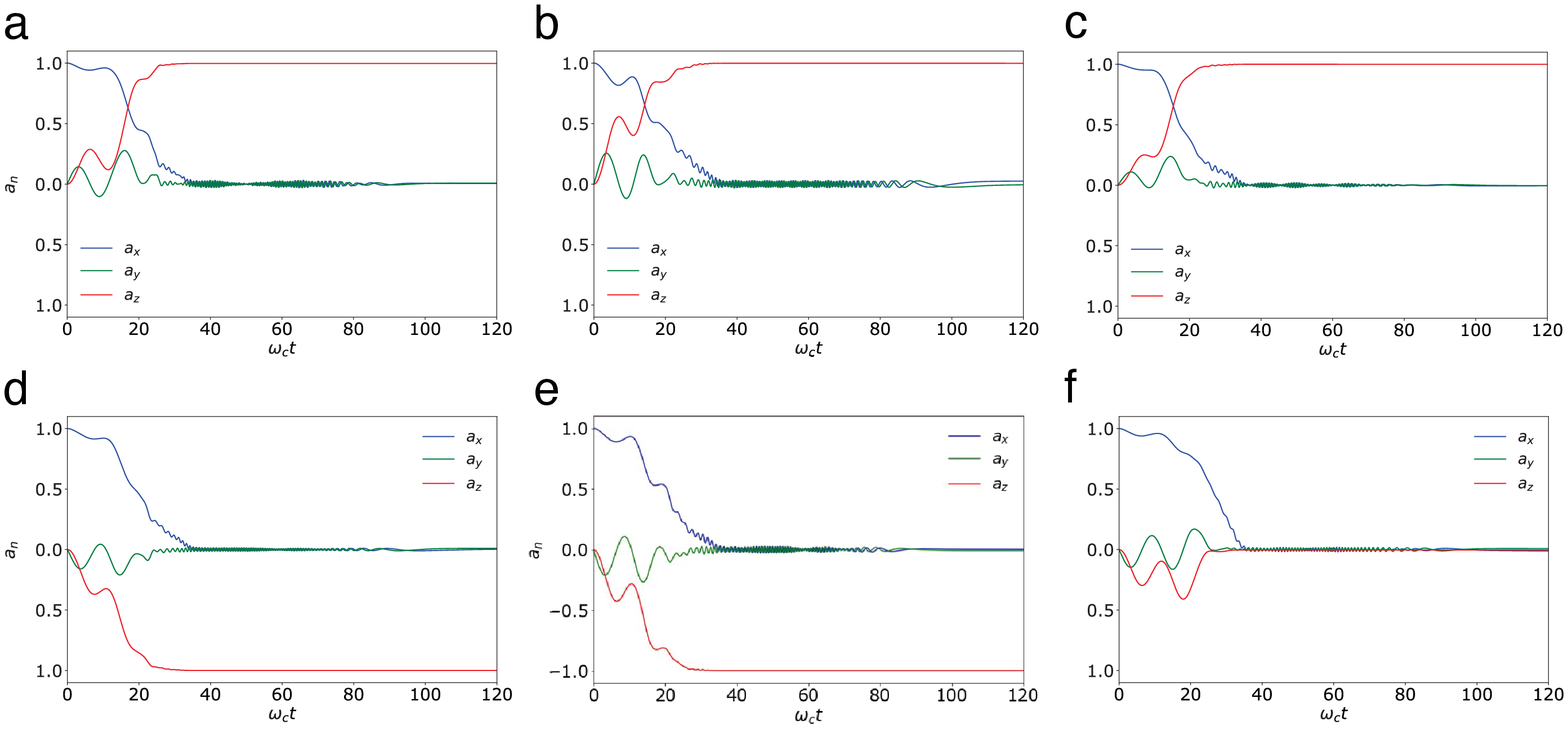}
\caption{Simulation of measurements of the polarization of the spin, with unbiased initial state $| \psi_{\rm{O}}(0) \rangle = |+ x \rangle$ of the spin and random initial conditions of meter and environment, Eqs.\ (\protect{\ref{rhopfunc}} - \protect{\ref{hightpfunc}}), modelled by the Hamiltonian (\protect{\ref{spinbosonham}}) with time-dependent modulations $f_{\rm O}(t)$ and  $f_{\rm OE}(t)$ as shown in Fig.\ \protect{\ref{fig6_modulasymp}}a. Panels (a) to (f) show the three components $a_x(t)$ (blue), $a_y(t)$ (green), and $a_z(t)$ (red) of the Bloch vector vs.\ time for different specific initial conditions of the environment. Time is scaled as $\omega_{\rm{c}} t$. The number of boson modes is $N = 300$, the Kondo parameter $\alpha$ varies with $f_{\rm OE}(t)$, reaching a maximum of $\alpha = 2$. Other parameter values are as in Fig.\ \protect{\ref{fig3_blochvstnomodul}},.
}
\label{fig5_blochvstmodul}
\end{center}
\end{figure}

\begin{figure}[h!]
\begin{center}
\includegraphics[width=8.6cm]{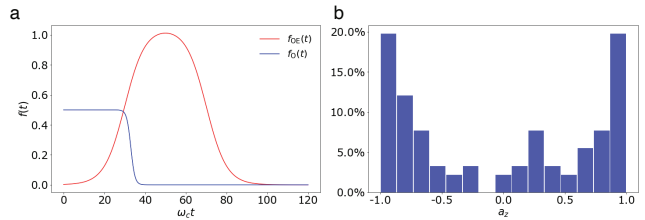}
\caption{(a) Modulation functions $f_{\rm O}(t)$ (blue) and $f_{\rm OE}(t)$ (orange) underlying the simulations shown in Fig.\ \protect{\ref{fig5_blochvstmodul}}, (b) distribution of long-time asymptotes $a_z(t)$ for $t \gg 2\pi / \omega_{\rm{c}}$, based on 100 simulations. Parameter values are as in Fig.\ \protect\ref{fig5_blochvstmodul}
}
\label{fig6_modulasymp}
\end{center}
\end{figure}

The sample simulations presented in Fig.\ \ref{fig5_blochvstmodul} have been obtained for the modulation functions shown in Fig.\ \ref{fig6_modulasymp}a. We now observe a strong tendency of the long-time averages of the Bloch vector to converge against either one of the poles of the Bloch sphere (Fig.\ \ref{fig5_blochvstmodul}a-e), approaching almost pure states close to eigenstates of the measured operator $\hat\sigma_z$. Only few initial conditions of the environment lead to persistently large fluctuations of $a_z(t)$ or to stable values close to $a_z(t) = 0$ (Fig.\ \ref{fig5_blochvstmodul}f). The $x$- and $y$-components decay to sustainedly very low levels. This behaviour is reflected in the statistics of long-time averages of $a_z(t)$, Fig.\ \ref{fig6_modulasymp}b, based on a total of 100 simulations (in contrast to Fig.\ \ref{fig4_asympolarnomodul}, no simulation has been discarded). The peaks close to the extremes  $a_z(t) = \pm 1$ are now more marked, the intermediate residual plateau is lower than in Fig.\ \ref{fig4_asympolarnomodul}.

We can compare these findings, for the time-independent as well a for the modulated version of the Hamiltonian, with previous studies of decoherence, based on non-unitary evolution equations for the reduced density operator of the spin coupled to an infinite heat bath. Analytical calculations in the context of quantum measurement \cite{Zur84,Zur03} indicate that the spin is projected towards the $z$-axis of the Bloch sphere, but without any preference for its poles. Detailed studies of the spin-boson model, using advanced numerical simulation techniques developed for molecular and solid-state physics \cite{WT08,WT10,KA13}, revealed a crossover from localization (attraction towards the $z$-axis of the Bloch sphere with non-zero polarization) to relaxation (attraction towards the origin of the Bloch sphere, i.e., complete depolarization), depending on specific features of the coupling to the heat bath and the nature of its frequency spectrum, sub-Ohmic, Ohmic, or super-Ohmic. A deeper analysis why in our case, with a finite heat bath, we do observe a preference for the poles will require further research.

\subsubsection{\label{sec332} Purity, partial, and mutual entropies}

The two conditions that after the measurement, the environment, representing meter and apparatus in our model, bears information on the state of the spin, and that the final state of the spin depends in turn on the initial state of the environment, suggest that as a consequence of their entanglement during the measurement, the two subsystems have bilaterally exchanged  information. This process should be reflected in related quantities, in particular in their purity and their partial and mutual entropies. With the simulations detailed above, we also monitored the time evolution of these quantities. The linear entropy
\begin{equation} \label{linent}
S_{\rm{O}}^{\rm{lin}}(t) = 1 -  {\rm{tr}}_{\rm{O}}\bigl([\bar{\hat\rho}_{\rm{O}}]^2\bigr)
\end{equation}
of the reduced density operator $\bar{\hat\rho}_{\rm{O}} = {\rm{tr}}_{\rm{E}} [\hat\rho]$ indicates how far the spin deviates from purity. Its entanglement with the apparatus is quantified by the partial entropy
\begin{equation} \label{partent}
S_{\rm{O}}(t) = -  {\rm{tr}}_{\rm{O}}(\bar{\hat\rho}_{\rm{O}}(t)) \ln\bigl({\rm{tr}}_{\rm{O}}[\bar{\hat\rho}_{\rm{O}}(t)]\bigr).
\end{equation}
The partial entropy $S_{\rm{E}}(t)$ of the environment is defined analogously. An equivalent measure of entanglement is the mutual or entanglement entropy \cite{EK95},
\begin{equation} \label{mutuent}
S_{{\rm{O}}\cap{\rm{E}}}(t) = S_{\rm{O}} + S_{\rm{E}} - S_{\rm{OE}},
\end{equation}
where $S_{\rm{OE}}$ denotes the entropy of the total system. For a bipartite system that as a whole is in a pure state as in our model, a Schmidt decomposition of the total state allows to show that the two partial entropies are identical \cite{EK95,CA97}, 
\begin{equation} \label{bipartent}
S_{\rm{O}}(t) = S_{\rm{E}}(t).
\end{equation}
Since in this case, the entropy of the total system vanishes, Eq.\ (\ref{mutuent}) implies that
\begin{equation} \label{mutubipartent}
S_{{\rm{O}}\cap{\rm{E}}}(t) = 2S_{\rm{O}} = 2S_{\rm{E}}.
\end{equation}

\begin{figure}[h!]
\begin{center}
\vspace{-0.5 cm}
\includegraphics[width=8.6cm]{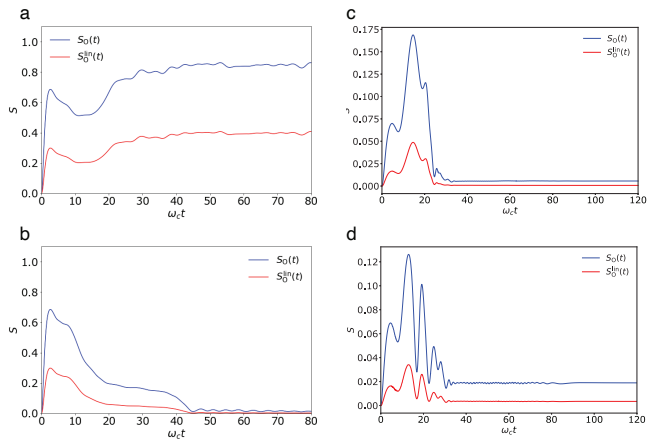}
\caption{Time evolution of the partial (blue) and the linear entropy (red) of the spin, with unbiased initial state $| \psi_{\rm{O}}(0) \rangle = |+ x \rangle$ of the spin and random initial conditions of meter and environment, Eqs.\ (\protect{\ref{rhopfunc}} - \protect{\ref{hightpfunc})}, modelled by the Hamiltonian (\protect{\ref{spinbosonham}}) with time-independent modulations $f_{\rm O}(t) = \rm{const}$ and  $f_{\rm OE}(t) = \rm{const}$ (panels a, b) and time-dependent modulations as shown in Fig.\ \protect{\ref{fig6_modulasymp}}a (panels c, d). Time is scaled as $\omega_{\rm{c}} t$, parameter values are as in Fig.\ \protect{\ref{fig3_blochvstnomodul}} (panels a, b) or Fig.\ \protect{\ref{fig5_blochvstmodul}} (panels c, d).
}
\label{fig7_entropurity}
\end{center}
\end{figure}

In Fig.\ \ref{fig7_entropurity}, we therefore compare the time evolution of the linear entropy (red) only with that of the partial entropy $S_{\rm{O}}$ of the spin (blue), for the two cases featured above, time-independent Hamiltonian as in Figs.\ \ref{fig3_blochvstnomodul}, \ref{fig4_asympolarnomodul} (panels a,b) and modulation of self-energy and coupling terms according to a standard measurement protocol as in Figs.\ \ref{fig5_blochvstmodul}, \ref{fig6_modulasymp} (panels c,d). With constant terms of the Hamiltonian, the scenario is as varied as it was found already for the underlying dynamics of the Bloch vector. From their initial value zero, the entropies rise to values close to 0.5, indicating strong entanglement of the two subsystems, but then show a behaviour ranging from localization in almost pure states of the spin to strong oscillations sustaining a high level of the entropies. For a time-dependent measurement protocol, the behaviour is much less ambiguous. An initial increase of the partial entropy to moderate values around 0.1 can be attributed to the first collapse of the wavefunction, reflected in considerable entanglement. Subsequently, upon switching off the coupling with the apparatus, it settles down to values of the order of 0.01, thus returning to a nearly pure state of the spin and in agreement with the almost complete localization of the Bloch vector. This scenario suggests being interpreted as a round trip from a pure initial state through an intermediate mixed phase back to a final state where both subsystems are close to purity, but after having  exchanged considerable information between them. Besides this main conclusion, a surprising observation is that even after decoupling the spin from the environment, in all simulations a positive residual mutual entropy remains that does not wane on the time scales covered, indicating that the measured spin never gets completely disentangled from the apparatus.

\subsubsection{\label{sec333} Redundant measurements and repeatability}

Besides the common scheme of measurements of $\hat\sigma_z$ with initial conditions of spin and apparatus that do not impose any systematic bias on the outcome, we also performed simulations where, as a consistency check, the spin is already in or close to an eigenstate of the measured observable upon entering the apparatus. The simplest instance of such redundant measurements is the case that the spin is initially in an eigenstate of $\hat\sigma_z$ and is stabilized there by a self-energy term $\sim \hat\sigma_z$ of the corresponding sign, instead of $\hat{H}_{\rm O} \sim \hat\sigma_x$ as in Eq.\ (\ref{oham}). The total Hamiltonian thus takes the form
\begin{align} \label{spinbosonhamz}
\hat{H} =& \frac{1}{2} \hbar\omega_0\hat\sigma_z f_{\rm O}(t) +
\sum_{n=1}^N  g_n \hat\sigma_z (\hat a_n^\dagger + \hat a_n)\, f_{\rm OE}(t) + \nonumber \\
&+ \sum_{n=1}^N \hbar\omega_n \left(\hat a_n^\dagger \hat a_n + \frac{1}{2}\right).
\end{align}
By contrast to the spin-boson model (\ref{spinbosonham}), this Hamiltonian commutes with $\hat\sigma_z$, so that it conserves eigenstates of this operator. It is therefore trivially guaranteed that after the measurement, the spin is still in the same eigenstate where it had been prepared.

Less obvious is the case of repeated measurements of the same observable. According to the projective measurement scheme, a second measurement on an object that on exit from the first run has been in an eigenstate of the measured observable, should yield again the same eigenvalue of this eigenstate. This postulate ensures the repeatability of quantum measurements \cite{NC00}. Within our model, we simulate repeated measurements with the following protocol:
\begin{enumerate}
\item  Prepare the spin in an unbiased initial state $| \psi_{\rm{O}}(0) \rangle = |+ x \rangle$ and the environment in a random initial condition, Eqs.\ (\protect{\ref{rhopfunc}} - \protect{\ref{hightpfunc})}, as for a common unbiased measurement.
\item Simulate a measurement of $\hat\sigma_z$ according to the Hamiltonian (\ref{spinbosonham}), with time-dependent modulations $f_{\rm O}(t)$ and  $f_{\rm OE}(t)$ as shown in Fig.\ \ref{fig6_modulasymp}a.
\item Assuming that after the measurement, the spin is already in a state close to $|-_z \rangle$ or $|-_z \rangle$, renormalize the Bloch vector $\mathbf{a} = (a_x,a_y,a_z)$ to $|\mathbf{a}|^2 = 1$, corresponding to a pure state.
\item Repeat the measurement of $\hat\sigma_z$ as above, but keeping the self-energy term (\ref{oham}) switched off and initiating the environment in a new random initial condition, independent of its final state on exit of the first measurement.
\end{enumerate}
Step 3 of this protocol, purifying the spin ``by hand'', may appear as an \emph{ad hoc} intervention. It can be justified heuristically, however, by the fact that before this renormalization, the spin states are already very close to purity, so that the remaining discrepancy we are removing can be attributed to deficiencies of our model and its implementation.

\begin{figure}[h!]
\begin{center}
\vspace{-0.5 cm}
\includegraphics[width=8.6cm]{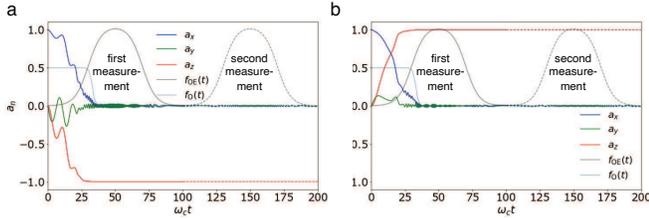}
\caption{Simulation of two subsequent measurements of the same observable $\hat\sigma_z$ of the spin, with unbiased initial state $| \psi_{\rm{O}}(0) \rangle = |+ x \rangle$ of the spin and random initial conditions of the environment, Eqs.\ (\protect{\ref{rhopfunc}} - \protect{\ref{hightpfunc}}) for the first measurement. The protocol, see text, is implemented based on the Hamiltonian (\protect{\ref{spinbosonham}}) with time-dependent modulations $f_{\rm O}(t)$ and  $f_{\rm OE}(t)$. The state of the spin after the first measurement (full lines) is kept as entrance condition of the second measurement (dashed lines), the environment is initiated in a new random condition, see step 4 of the protocol. The figure shows the components of the Bloch vector (colour code as in Fig.\ \protect\ref{fig3_blochvstnomodul})  vs.\ time for a first measurement resulting in the eigenstates $| -_z \rangle$ (panel a) and $| +_z \rangle$ (b). Light blue and grey lines indicate the modulation functions $f_{\rm O}(t)$ and $f_{\rm OE}(t)$, resp. Time is scaled as $\omega_{\rm{c}} t$, parameter values are as in Fig.\ \protect{\ref{fig5_blochvstmodul}}.
}
\label{fig8_redumeas}
\end{center}
\end{figure}

In Fig.\ \ref{fig8_redumeas}, we depict the time evolution of the Bloch vector for two such repeated measurements, one where the result of the first run has been spin down (panel a), the other where it has been spin up (b). The principal observation in all these simulations is that the second measurement faithfully reproduces the result of the first one. We thus conclude that, as concerns repeatability, our model satisfies the postulates of projective measurement.

\subsubsection{\label{sec334} Linear superpositions of initial states}

An important critical argument in the context of a unitary approach to quantum measurement refers to its consistency with a basic tenet of quantum mechanics, the principle of linear superposition. In essence, in the case of spin measurement, it reads as follows: Consider two initial states of the total system of the form as in Eqs.\ (\ref{oeini},\ref{rhoeini}), which in the long-time limit converge against opposite eigenstates of the measured operator $\hat\sigma_z$, say
\begin{equation} \label{pmini}
\begin{split}
| \Psi_{0-} \rangle &=  |-_x \rangle | \psi_{{\rm{E}}0-} \rangle \xrightarrow[t \to \infty]{} |-_z \rangle | \psi_{{\rm{E}}\infty -} \rangle, \\
| \Psi_{0+} \rangle &= |-_x \rangle | \psi_{{\rm{E}}0+} \rangle \xrightarrow[t \to \infty]{} |+_z \rangle | \psi_{{\rm{E}}\infty +} \rangle.
\end{split}
\end{equation}
Can we expect that a coherent superposition of these initial states,
\begin{align} \label{phiini}
| \Psi_0(\phi) \rangle &= \cos(\phi) | \Psi_{0-} \rangle +  \sin(\phi) | \Psi_{0+} \rangle \nonumber\\
&= |-_x \rangle \bigl(\cos(\phi) | \psi_{{\rm{E}}0-} \rangle + \sin(\phi) | \psi_{{\rm{E}}0+} \rangle \bigr),
\end{align}
with $\phi \in [0,\pi/2]$, will also evolve towards a definite outcome, spin up or down? The mixing angle $\phi$ parameterizes the Hilbert space spanned by the states $| \Psi_{0-} \rangle$ and $| \Psi_{0+} \rangle$. Varying it from $\phi = 0$ to $\phi = \pi/2$, the state $| \Psi_0(\phi) \rangle$ scans the continuous range of initial conditions from $| \Psi_0(0) \rangle = | \Psi_{0-} \rangle $ through the symmetric superposition
\begin{equation} \label{stoicaini}
| \Psi_0(\pi/4) \rangle = \frac{1}{\sqrt{2}} \bigl( | \Psi_{0-} \rangle +  | \Psi_{0+} \rangle \bigr)
\end{equation}
through $| \Psi_{0}(\pi/2) \rangle = | \Psi_+ \rangle $. For the symmetric case (\ref{stoicaini}), the answer should be negative, since a definite outcome would break the parity $z \to -z$ of that initial state.

Checking this question experimentally would not have been possible with the simple Davydov ansatz (\ref{d2ansatz}), since a linear superposition of coherent states is not a coherent state in itself, as required for this scheme. However, the multimode Davydov ansatz (\ref{d2multi}) is sufficiently general to include also superpositions of pairs of initial states of the environment, specifically those which in previous simulations have led to opposite spin eigenstates. It only requires to augment the original multiplicity $M$ of these states to $2M$ for their superposition.

\begin{figure}[h!]
\begin{center}
\includegraphics[width=8.6cm]{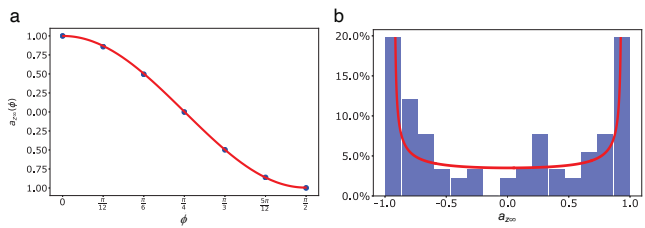}
\caption{Simulation of measurements with the environment prepared in linear superpositions, (Eq.\ \protect\ref{phiini}), of pairs of initial states that converge to opposite outcomes, parameterized by a mixing angle $\phi$. (a) Long-time averages $a_{z\infty}(\phi)$ of the polarization for discrete values $\phi_n = n \pi/12$, $n = 0,\,\ldots,\,6$ (blue dots), compared to the fit (\protect\ref{azfinphi}) (red curve). (b) Probability density, Eq.\ (\protect\ref{probazfin}) (red curve), corresponding to the fit (\protect\ref{azfinphi}), compared to the histogram (blue) of $a_{z\infty}$ shown in Fig.\ \protect\ref{fig6_modulasymp}b.
}
\label{fig9_stoica}
\end{center}
\end{figure}

Figure \ref{fig9_stoica}a shows the results of such simulations, where the mixing angle has been varied in six discrete steps between $\phi = 0$ and $\phi = \pi/2$, in terms of the long-time average $a_{z\infty}(\phi)$ of the $z$-component of the Bloch vector (blue dots). As expected, for $\phi = \pi/4$ the reduced density of the spin converges to a symmetric depolarized state with $|\mathbf{a}|^2 = 0$. For all biased superpositions $\phi \neq \pi/4$, however, it tends to localize at states with $a_{z\infty} \neq 0$. In particular, the transition from $a_{z\infty}(0) \approx 1$ to $a_{z\infty}(\pi/2) \approx -1$ is not linear but occurs predominantly in the range of angles close to $\pi/4$, giving strong weight to the marginal values $a_{z\infty} \approx \pm 1$.

If we assume an {\em{incoherent}} superposition of the long-time asymptotic states of the environment, with probabilities $p(|\Psi_{\infty +} \rangle) = \bigl(\cos(\phi)\bigr)^2$ and $p(|\Psi_{\infty -} \rangle) = \bigl(\sin(\phi)\bigr)^2$, resp., we would expect a projective quantum measurement to result in a measured polarization (red continuous curve in Fig.\ \ref{fig9_stoica}a)
\begin{equation} \label{azfinphi}
a_{z\infty}(\phi) = (+1) \bigl(\cos(\phi)\bigr)^2 + (-1)\bigl(\sin(\phi)\bigr)^2 = \cos(2\phi).
\end{equation}
It is the same outcome projective measurement would predict for initial states $| \Psi_0(\phi) \rangle = \bigl(|+_z \rangle \cos(\phi) + |-_z \rangle \sin(\phi)  \bigr) | \psi_{{\rm{E}}0} \rangle$ instead of Eq.\ (\ref{phiini}), with unbiased random environment state $| \psi_{{\rm{E}}0} \rangle$. This fit allows to estimate the probability density distribution of $a_{z\infty}(\phi)$ as 
\begin{equation} \label{probazfin}
p(a_{z\infty}) \sim \sqrt{1 + 1 / (1 - a_{z\infty}^2)}
\end{equation}
Comparing Eq.\ (\ref{probazfin}) (red curve) in Fig.\ \ref{fig9_stoica}b with the histogram of long-time averages shown above in Fig.\ \ref{fig6_modulasymp}b, we find remarkable agreement. In this sense, our numerical results are consistent with the principle of linear superposition. This would suggests the uncomfortable conclusion that the nonzero probability of indefinite outcomes near $a_{z\infty} = 0$ is an unavoidable feature for a unitary description of spin measurement including the object and a macroscopic environment.

\section{\label{sec4} Conclusions}

In this work, we explored simulations of spin measurements, following a hybrid approach that combines a unitary account of  measured object and apparatus with a finite heat bath model of the environment, thus avoiding a statistical description of the state of the spin. Realistic measurement protocols, separating preparation, measurement, and post-measurement phases, could be implemented by a time-dependent modulation of self-energy and coupling terms in the Hamiltonian. With a moderate number of a few hundred boson modes and a frequency spectrum adopting spectral densities common in quantum optics and solid-state physics, we have been able to approximately reproduce all essential features of projective measurements, the initial entanglement of the measured object with meter and apparatus leading to the (first) collapse of the wavefunction as well as their subsequent disentanglement and the approach of the spin to one of the eigenstates of the measured observable, corresponding to the second collapse. In the case of an unbiased initial preparation of the spin, we could simulate the stochastic nature of this long-time asymptote, equivalent to the result of the measurement, and trace it back trial by trial to the random initial condition of the environment that lead to this outcome. In the light of our unitary approach, quantum randomness of the measurement thus appears as a manifestation of the thermal randomness of a macroscopic many-body system. In complementary simulations of iterated measurements of the same observable, we could verify the repeatability of the first result.

To be sure, the basic features of projective quantum measurements have only been met in an approximate sense. In our simulations, the first collapse never leads to a complete loss of coherence, the second collapse never to a perfect restoration of a pure state of the spin. In particular, the long-time averages of the measured observable, the polarization of the spin, do not satisfy the exclusive discrete alternative, either spin up or spin down with 50\% probability each. Rather they show a smooth probability distribution with a significant residual plateau between these extremes. However, in view of the history of the theory of quantum measurement, this attenuation does not come as a surprise: For example, the detailed microscopic modelling of the first collapse \cite{Zur81,JZ85,Ven97} replaced the discontinuous jump from a coherent superposition to an incoherent sum, postulated by the traditional account, by a gradual decay taking place in finite time, epitomized in the headline ``Collapse of the wavepacket: how long does it take?'' \cite{Zur84}, and never ending in perfect decoherence.

Known analytical and numerical results for the spin-boson model with a continuous spectrum even indicate that the spin is completely depolarized or at most attracted to the $z$-axis of the Bloch sphere \cite{Ven97,Zur03,WT08,WT10,KA13} but not to its poles. Based on these findings, we could not expect either to find the preference for these points in the projective Hilbert space. In fact, our numerical simulations of the superposition argument \cite{Sto16} show that the unitary approach to quantum measurement is compatible with the principle of linear superposition, if and only if we replace the exclusive alternative by a smooth continuous distribution that includes a nonzero, if small, probability for intermediate, in particular indefinite, outcomes of the measurement. In addition this means that our simulations do not generally lead to perfectly distinguishable, i.e., mutually orthogonal, states of a meter that could be interpreted as readouts. We are confident that a refined modelling, particularly treating meter and apparatus as separate subsystems \cite{VKG95} instead of merging them into an all-inclusive environment as we did here, and improved implementations of the numerical simulations will bring this approach even closer to a comprehensive description of quantum measurement.

We have not yet been able to simulate a situation that could serve as another touchstone of a unitary account of quantum measurement: Certain steps in the implementation of the Davydov ansatz that involve an inherent irreversibility, specifically the elimination of basis states in case of overlapping coherent states (``apoptosis'' \cite{WG20}), prevent simulating the measurement in inverse time, from the final outcome back to a neutral spin state and the specific initial condition of the environment. It is left as a task for future work. 

Our numerical simulations appear to be far away from an experimental realization. Yet this could be a realistic perspective. Indeed, for example, Raimond, Brune, and Haroche \cite{RBH97} proposed a detailed setup of a quantum optics experiment where in a pair of weakly coupled cavities, one of them serving as a ``single-mode reservoir'' but forming a closed system with the other, as in our model, decoherence is reached on a short time scale. On a longer time scale, the unitary nature of their common time evolution becomes manifest in periodic revivals. In a similar vein, it has been demonstrated that seemingly discontinuous processes induced by a macroscopic environment, such as quantum jumps, can not only be monitored as gradual transitions in continuous time but even be reversed ``mid-flight'' \cite{MM&19}, thus substantiating the compatibility of unitary accounts with apparent decoherence.

With our model, we have deliberately restricted ourselves to a setup consisting of a measurement on a single spin only. It is tempting to speculate about a similar account of experiments with pairs of correlated two-state systems \cite{Zei99}, observed in measurements at space-like separated sites, i.e., in EPR setups: In this context, quantum randomness is inextricably tied to quantum nonlocality, rising additional questions concerning a microscopic modelling of the pair of measuring apparatuses. Analyzing quantum randomness in EPR experiments in a correspondingly extended model, possibly including two separate heat baths, will be a challenge for future research.

As concerns the environment, we adopted the conventional model of a set of harmonic oscillators linearly coupled to the central system. The way entanglement and information are distributed in the environment and shared with the measured object is essential for the measurement process. Recent research indicates that closed but strongly interacting, possibly chaotic, quantum many-body systems do not only relax, but generally even approach thermal equilibrium \cite{MSS16,LB17,RUR18,Swi18}. Another question worth being explored is therefore in how far quantum chaos in the degrees of freedom of meter and apparatus would affect the exchange of information with the observed system and become manifest in the measurement process and the quantum randomness of its results.

The two-state system representing spins in our model can be considered as a low-energy approximation of the infinite-dimensional Hilbert space of a symmetric double-well system \cite{LC&87}. In a recent publication closely related to our work, Choudhury and Grossmann \cite{CG22} studied such a double-well potential in a similar quantum mechanical approach with finite heat bath as followed here. They find that the central system, if prepared initially in a symmetric state on the top of the barrier with the environment in the ground states of all boson modes, does not approach either one of the two potential minima but relaxes to its own symmetric ground state. Simulations with the environment prepared instead in a random initial state that could lead to a spontaneous symmetry breaking in the long-time behaviour are pending. An analogous spontaneous symmetry breaking, induced by the initial conditions of a finite heat bath, has indeed been found in a classical-mechanics account of a quartic double well coupled to a reservoir of harmonic oscillators \cite{DP20}.

\section*{Acknowledgements}
We enjoyed fruitful and inspiring discussions with Konstantin Beyer, Sreeja Loho Choudhury, Frank Grossmann, Walter Strunz, and Michael Werther at the Institute for Theoretical Physics, Technical University of Dresden (TUD). OR and TD thank for the hospitality extended to them by this Institute during various research stays where part of the work reported here has been performed. OR acknowledges financial support by \emph{Fundaci\'on para la promoci\'on de la investigaci\'on y la tecnolog\'\i a} (FPIT) of \emph{Banco de la Rep\'ublica de Colombia} (project 4050). TD further gratefully acknowledges free online access, granted to him by the Max Planck Institute for the Physics of Complex Systems (MPIPKS, Dresden, Germany) and by TUD, to their electronic journal libraries. He would like to dedicate this work to the late Fritz Haake, his academic teacher from whom he learned most of what he knows about quantum measurement.




%

\end{document}